\documentclass{aastex61}

\begin{document}
\title{Dynamics evolution of a solar active-region filament from quasi-static state to eruption: rolling motion, untwisting motion, material transfer, and chirality}
\email{yanxl@ynao.ac.cn}
\author{Xiaoli Yan}
\affiliation{Yunnan Observatories, Chinese Academy of Sciences, Kunming 650216, People's Republic of China.}
\affiliation{Key laboratory of Dark Matter and Space Astronomy, Chinese Academy of Sciences.}
\affiliation{Center for Astronomical Mega-Science, Chinese Academy of Sciences, 20A Datun Road, Chaoyang District, Beijing, 100012, People's Republic of China.}

\author{Qiaoling Li}
\affiliation{Yunnan Observatories, Chinese Academy of Sciences, Kunming 650216, People's Republic of China.}
\affiliation{University of Chinese Academy of Sciences, Yuquan Road, Shijingshan Block, Beijing 100049, People’s Republic of China.}

\author{Guorong Chen}
\affiliation{Yunnan Observatories, Chinese Academy of Sciences, Kunming 650216, People's Republic of China.}
\affiliation{University of Chinese Academy of Sciences, Yuquan Road, Shijingshan Block, Beijing 100049, People’s Republic of China.}

\author{Zhike Xue}
\affiliation{Yunnan Observatories, Chinese Academy of Sciences, Kunming 650216, People's Republic of China.}
\affiliation{Center for Astronomical Mega-Science, Chinese Academy of Sciences, 20A Datun Road, Chaoyang District, Beijing, 100012, People's Republic of China.}

\author{Li Feng}
\affiliation{Key laboratory of Dark Matter and Space Astronomy, Chinese Academy of Sciences.}

\author{Jincheng Wang}
\affiliation{Yunnan Observatories, Chinese Academy of Sciences, Kunming 650216, People's Republic of China.}
\affiliation{Center for Astronomical Mega-Science, Chinese Academy of Sciences, 20A Datun Road, Chaoyang District, Beijing, 100012, People's Republic of China.}

\author{Liheng Yang}
\affiliation{Yunnan Observatories, Chinese Academy of Sciences, Kunming 650216, People's Republic of China.}
\affiliation{Center for Astronomical Mega-Science, Chinese Academy of Sciences, 20A Datun Road, Chaoyang District, Beijing, 100012, People's Republic of China.}

\author{Yan Zhang}
\affiliation{Yunnan Observatories, Chinese Academy of Sciences, Kunming 650216, People's Republic of China.}
\affiliation{University of Chinese Academy of Sciences, Yuquan Road, Shijingshan Block, Beijing 100049, People’s Republic of China.}

\begin{abstract}
To better understand magnetic structure and eruptive process of solar filaments, a solar active-region filament (labeled F2) eruption associated with a B-class flare was investigated by using high-resolution H$\alpha$ data from the 1 m New Vacuum Solar Telescope (NVST), combined with EUV observations of the Solar Dynamical Observatory (SDO). The filament F2 was disturbed by another filament (labeled F1) eruption that experienced a whip-like motion. Before the filament F2 eruption, the Dopplergrams show that the southern and the northern parts of the filament F2 body exhibit blueshift and redshift along the filament spine, simultaneously. It implies that the filament F2 was rolling from one side to the other. During the filament F2 eruption, the Doppler velocity shifts of the filament body are opposite to that before its eruption. It demonstrates that the filament body exhibits an untwisting motion, which can be also identified by tracing the movement of the eruptive filament threads. Moreover, it is found that the material of the filament F2 was transferred to the surrounding magnetic field loops, which is caused by magnetic reconnection between the filament F2 and the surrounding magnetic loops. According to the right-bearing threads of the filament F2 before its eruption, it can be deduced that the filament F2 is initially supported by a sheared arcade. The following observations reveal that the twisted magnetic structure of the filament F2 formed in the eruption phase.
\end{abstract}


\section{Introduction}\label{sec:introduction}

Solar filaments/prominences suspended in the solar corona are large-scale magnetic structures with relatively cool and dense materials (Tandberg-Hanssen 1995). The evolution of solar filament number with time in one solar cycle is very similar to that of sunspots (Kong et al. 2014). According to the locations of solar filaments, they are classified into quiescent filaments, active-region filaments, and intermediate filaments (Engvold 1998). When solar filaments are born in the active regions, they are called active-region filaments. Compared with quiescent filaments, active-region filaments are relatively short-lived structures lasting for several hours or days (see the reviews of Parenti 2014 and Vial \& Engvold 2015). Besides, observations confirm that most of solar filaments in the northern hemisphere have negative helicity and those in the southern hemisphere have positive helicity (Martin 1998; Pevtsov et al. 2003).  By using the method developed by Chen et al. (2014), Ouyang et al. (2017) further confirmed this hemispheric rule of solar filament helicity. 

There are two popular points of view on the magnetic structure of solar filaments (see the review of Mackay et al. 2010). One is the sheared arcade structure (Kippenhahn \& Schl{\"u}ter 1957; Antiochos et al. 1994; DeVore \& Antiochos 2000). The other is the twisted flux rope structure (Kuperus \& Raadu 1974; van Ballegooijen \& Martens 1989; Amari et al. 2000; Gibson \& Fan 2006). The former magnetic structure indicates a normal-polarity configuration, which means that the magnetic field lines of solar filament are directed from positive polarity to negative polarity across the polarity inversion line (PIL). In this model, the filament material is suspended in the dip of sheared arcade structure and magnetic tension supports the material against its gravity. The latter magnetic structure indicates an inverse-polarity configuration, which means that the magnetic field lines of solar filament are directed from negative polarity to positive polarity across the filament. In this case, the filament material is located at the bottom of the helical magnetic field lines. The magnetic tension force provides the support of the filament material against its gravity. Guo et al. (2010) found that the magnetic field structure in one part of a filament is a flux rope, while the one in the other part is a sheared arcade. 

In fact, it is very difficult to identify whether solar filaments have sheared arcade or flux rope magnetic structures before their eruptions. Usually, the dark structure of solar filaments is the projection of the filament material on the solar disk. The real 3-D magnetic structure of solar filaments cannot be obtained by direction observations from quasi-static state evolution due to lack of direct measurement of filament magnetic field. However, once a filament is triggered to erupt, the twisting motion can be identified in the Doppler observations. Schmieder, Raadu, \& Malherbe (1985) found a twisting motion in a disturbed solar filament during its expansion. By using Huairou H$\beta$ dopplergram observation, Wang et al. (1996) reported a disturbing filament with a pigtail structure after magnetic cancellation under the filament in the photosphere. Panasenco, Martin, \& Joshi (2011) found that if a filament erupts non-radially, as frequently happens, the top of its spine first bends to one side and evolves into a sideways rolling motion. Using the NVST data, Awasthi et al. (2019) reported that there are two kinds of mass motion in an active-region filament, i.e., rotation about the spine and longitudinal oscillation along the spine driven by the disturbance of a surge. Besides, untwisting motion of solar filaments is also found during their eruptions (Kurokawa et al. 1987; Yan et al. 2014a, 2014b; Yang et al. 2014; Chen et al. 2019). All observations imply that these solar filaments may have twisted magnetic structure. Besides the imaging observation, several authors found the existence of a twisted magnetic structure at the location of the filament by using non-linear force free extrapolation (Feng et al. 2013; Jiang et al. 2014; Wang et al. 2015; Yan et al. 2015). 

Up to now, it is still unclear whether the twisted structure of solar filaments forms before or during their eruptions. Yan et al. (2015) suggested that shearing motion and sunspot rotation play an important role in the twisted magnetic structure formation of two active-region solar filaments before their eruptions. Sunspot rotation was also found in the formation of a solar filament by Chen et al. (2018). Pant et al. (2018) reported that the counterclockwise rotation of the two footpoints results in the opposite twist at the spine of the prominence, leading to prominence eruption. Furthermore, magnetic reconnection is also observed during the formation of filament channels and filaments (van Ballegooijen \& Martens 1989; Schmieder et al. 2004; Wang \& Muglach, 2007; Yan et al. 2016; Xue et al. 2017; Wang et al. 2017; Yang \& Chen 2019). Some simulations also reveal that prominence plasma is supported in magnetic dips of the twisted flux rope (Xia \& Keppens 2016). There are some investigations indicating that the filaments are supported by a sheared arcade magnetic structure before their eruptions (DeVore \& Antiochos 2000; Ouyang et al. (2015, 2017). 

Solar filament eruptions are usually associated with solar flares and coronal mass ejections (CMEs), which can be explained by the standard flare/CME model (Vr{\v{s}}nak et al. 1987; Gilbert et al. 2000; Zhang, \& Wang 2001; Gopalswamy et al. 2003; Chen 2011; Shen et al. 2011; Sterling et al. 2018; Yang et al. 2018; Li et al. 2018; Li et al. 2019; Zou et al. 2019; Hou et al. 2020). However, in individual events, there are always some special features that are not covered in the standard model. Sometimes, solar filaments also experience failed eruptions without CMEs (Ji et al. 2003; Song et al. 2018). Alexander, Liu, \& Gilbert (2006) studied the failed filament eruption reported by Ji et al. (2003) and found that there were two hard X-ray sources during the filament eruption. They argued that the second source of coronal hard X-ray emission (the lower one) was caused by the interaction of the two adjacent legs underneath the writhing filament, which is also confirmed by the simulation of Kliem et al. (2010). Liu et al. (2008) studied a partial filament eruption on 2007 March 2 and suggested that the kink instability and internal tether-cutting may work together in initiating the filament eruption. Recently, Cheng, Kliem, \& Ding (2018) investigated a failed and a successful filament eruptions and found both the cases are partial eruptions due to internal reconnection of the filament-hosting magnetic flux ropes. Yan et al. (2020) reported the possible evidence on the external magnetic reconnection between the filament and the overlying magnetic loops during a failed filament eruption. The simulation of Zhang et al. (2020) revealed that feeding axial flux into a pre-existing flux rope is also a viable mechanism to cause the eruption of solar magnetic flux rope. Note that the detailed definition on filament mass and structure eruption types can be seen from Gilbert, Alexander, \& Liu  (2007). Therefore, different cases exhibit different eruption processes. With high resolution data being available, further research is needed to figure out the detailed filament eruption.

In this paper, we present two active-region filament eruptions associated with two B-class flares, focusing on the fine structures of the filament revealed by the New Vacuum Solar Telescope (NVST) and SDO observations. The paper is organized as follows: The observations are described in Section \ref{sec:observations}. The details of results are presented in Section \ref{sec:results}. The conclusion and discussion are given in Section \ref{sec:conclusion}.

\section{Observations}\label{sec:observations}
The 1-m NVST is designed to observe the fine structures of the photosphere and chromosphere of the Sun, which is located in the Fuxian Solar Observatory of the Yunnan Observatories, Chinese Academy of Sciences (CAS). The H$\alpha$ line-center, and the H$\alpha$ off-band images acquired at 6562.8~{\AA} and 6562.8 $\pm$ 0.4~{\AA} are employed to demonstrate the process of the two filament eruptions and to construct Dopplergrams. The spatial resolution is 0.$^\prime$$^\prime$3 and the temporal resolution is $\sim$12 s (Liu et al. 2014; Yan et al. 2019). The data were calibrated from Level 0 to Level 1 with the dark current subtracted and flat-field corrected. Furthermore, the calibrated images were reconstructed by using the speckle masking method from Level 1 to Level 1+ as done by Weigelt (1977), Lohmann et al. (1983), and Xiang et al. (2016). Furthermore, the Dopplergrams are constructed by using the following equation (Langangen et al. 2008), 
\begin{equation}
D = \frac{B - R}{B + R}
\end{equation}
where $B$ and $R$ represent the blue-wing (H$\alpha$ -0.4 \AA) and the red-wing (H$\alpha$ + 0.4 \AA) images, respectively. Note that the Dopplergrams just show the Doppler shift signals (not the exact Doppler velocity).

Solar Dynamic Observatory (SDO; Scherrer et al. 2012) is launched in 2010 September. It can provide multi-band EUV images and magnetograms. The EUV images obtained by the Atmospheric Imaging Assembly (AIA; Lemen et al. 2012) on board SDO, including the 304~\AA\, 171~\AA\, and 131~\AA\ channels, with a spatial resolution of $1.^{\prime\prime}5$ and a cadence of 12 s, are used to display the process of the filament eruption and the associated magnetic reconnection in the corona. Photospheric line-of-sight magnetograms (LOS) with a 45 s cadence and a pixel scale of 0.5$^\prime$$^\prime$ are provided by the Helioseismic and Magnetic Imager (HMI;  Schou et al. 2012) on board SDO to show the magnetic fields surrounding the two filaments.

\section{Results}\label{sec:results}
\subsection{Appearance of active region NOAA 12740}
Active region NOAA 12740 appeared at the east solar limb on 2019 May 5. The original magnetic configuration was of $\alpha$-type. On 2019 May 6, the magnetic configuration became $\beta$$\delta$-type. Following the complexity of the magnetic configuration, eruptive activities became more and more frequent. The active region on 2019 May 7 is shown in Figure 1. This active region is located at the northern hemisphere of the Sun near the solar equator. The biggest sunspot is the leading one with negative polarity (see Figures 1(a) and 1(b)). The following positive polarity is disperse. There are some small positive polarities around the leading sunspot (see Figure 1(b)). Figures 1(c) and 1(d) show the EUV observations (304 \AA\ and 131 \AA\ images) of this active region with the line-of-sight magnetogram superimposed. The levels of the magnetic contour are $\pm$ 200 G. The yellow curved lines in Figure 1(c) and 1(d) indicate the location of an active region filament labeled F2 in the following text. Since the filament labeled F1 in the following text cannot be clearly seen in Figure 1(b) and 1(c), it is not marked in two figures. Seen from Figure 1(c) and 1(d), one end of the filament is rooted in the leading sunspot with negative polarity and the other end is rooted in the small positive polarity near the leading sunspot. 

\subsection{The filament F2 eruption triggered by the filament F1 eruption}
On May 7, 2019, the NVST observed the eruptions of two S-shaped active-region filaments in active region (AR) NOAA 12740. Figure 2 shows the eruption process of the two filaments in the H$\alpha$ center images (see animation 1). The line-of-sight magnetic fields were superimposed on the H$\alpha$ center image (see Figure 2(a)). The contour values of the magnetic fields are $\pm$ 100 G. The red and the blue contours denote the negative and the positive polarities. One of the two filaments labeled F1 is located at the northwest of the leading sunspot (see the white arrows in Figs. 2(a) and 2(c)). The other filament labeled F2 is located at the northeast of the leading sunspot (see the yellow arrows in Figs. 2(a)-2(c)). There are also some scatter positive polarities near the north of the leading sunspot. By tracing the threads of two filaments, it is found that one foot-point of each filament was rooted in the penumbra of the leading sunspot. The other foot-point of each filament was rooted in the surrounding weak positive polarity. Moreover, both filaments exhibited an S-shaped configuration. 

At about 09:16 UT, the filament F1 was activated and then erupted. An associated B5.0 solar flare started at 09:18 UT, peaked at 09:23 UT, and ended at 09:25 UT according to the GOES flare record. The filament F1 eruption looks like a whip-like motion with the northern part erupting and the southern end being almost fixed (see the white arrow in Figure 2(c)). In the following, the filament F1 erupted along the direction of the large-scale sunspot super-penumbral fibrils. From SDO/AIA EUV observations, the material of the filament F1 was transferred into the large-scale magnetic loops connecting the positive polarity at the middle of the filament F2. 

According to the H$\alpha$ observations, some threads at the eastern end of the filament F2 began to erupt from closed to open during the period  from about 05:28 UT to 06:20 UT. The similar process can be also seen in the H$\alpha$ observations from 08:35 UT to 09:15 UT. After the filament F1 eruption, there is an elongated brightening (ribbon-like) to the north of F2 at 09:23 UT (see the blue arrow in Fig. 2(b)). Such a brightening concurrent with the eruption of F1 is strongly indicative of magnetic connectivity between the magnetic systems of the two filaments. From the observations of SDO/AIA, one footpoint of a group of magnetic loops overlying the filament F1 is located in the brightening region. During the F1 eruption, the F2 was activated completely at about 09:30 UT. From a series of the H$\alpha$ center images, the arched structure of the F2 can be seen by tracing its threads marked by the red arrows in Figure 2(d) before the eruption. The obvious brightening appeared around the F2 (see Figure 2(e)). According to the GOES flare record, an associated B8.0 solar flare started at 09:29 UT, peaked at 09:35 UT, and ended at 09:39 UT. After the F2 eruption, an obvious untwisting motion of the F2 body was observed by tracing the eruptive threads of the filament. The filament body rotates clockwise if viewed from the top to solar surface. Moreover, the twisted structure can be seen from the evolution of the F2 threads (see the red arrows in Figure 2(f)-2(i)). The twist of the filament was estimated to be at least one turn by tracing the threads winding of the filament. According to the observation mentioned above, it is determined that the filament has positive helicity.

\subsection{Evolution of Doppler shift of the filaments F1 and F2 before and after its eruption.}
The H$\alpha$ off-band $\pm$ 0.4 \AA\ images are used to construct the Dopplergrams. Figure 3 shows the Dopplergrams before and during the F1 and F2 eruptions (see animation 2). The blue and the red arrows denote the Doppler blueshift and redshift. Before the F2 eruption, its Doppler map can be divided into two parts. The southern part exhibits blueshift and the northern part exhibits redshift (see the blue and the red arrows in Figures 3(a1)-3(a3)). Especially, this phenomenon in the east part of the filament is much clearer than that in the west part of the filament. It implies that the filament was rolling from the southern side to the northern side. The rolling motion gradually became weak close to the onset of the filament eruption (see Figures 3(a4)-3(a7)). At the beginning stage of the filament eruption, the blue and the red Doppler shifts appeared alternately along the spine of the filament axis (see Figure 3(a8)). It should be noted that the red and the blue shifts should be reversed when the H$\alpha$ line becomes emissive. Since several flare kernels are visible at this moment due to the emission of the H$\alpha$ line in Figure 3(a8), we reverse the signs of Doppler velocity at the emissive region. During the F2 eruption, the southern part exhibits redshift and the northern part exhibits blueshift (see Figures 3(a9)-3(a12)). The Doppler shift is completely opposite to that before the filament eruption. It reveals that the filament began to experience untwisting motion. These observations also indicate that the magnetic structure of the filament became a twisted flux rope. The onset of F1 eruption was prior to that of the filament F2 eruption. The duration of the filament F1 eruption lasted from about 09:16 UT to 10:00 UT. At the original stage of F1 eruption, the eruptive structure exhibited blueshift (see Figs. 3(a5)-3(a7)). In the following, the Doppler shift along the filament F1 eruption became redshift (see Figs. 3(a8) - 3(a12)). It implies that a part of the erupting material fell down along the super-penumbra fibrils of the leading sunspot. 

\subsection{Materials transfer during the filament F2 eruption}
The white and the black arrows in Figure 4(a1), 4(b1), and 4(c1) denote the locations of the filaments F1 and F2. At the beginning of F1 eruption, the obvious brightening appeared at the site of F1 (see the white arrows in Figs. 4(a1), 4(b1), and 4(c1)). After the F1 eruption, its material was transferred into the large-scale magnetic loops (see the white arrows in Figs. 4(a2), 4(b2), and 4(c2)). These loops connected from the leading sunspot to the middle of the filament F2. The trajectory of the filament material transfer is outlined by a dotted red line in Figure 4(c2). When the material was injected into these loops, the brightening appeared at the foot-points of these loops (see the northern end of the dotted line in Fig. 4(c2) and the animation 3 at about 09:22 UT in 131 \AA\ observations). Even though the ejected material from the filament F1 is still half way from the eruption site, the large-scale magnetic loops connecting the middle of the filament F2 can be seen to be heated due to the heat conduction from the site of the filament F1 eruption (see the time period from 09:21: 30 UT to 09:24: 54 UT in the animation 3). Moreover, the large-scale magnetic loops expand toward the northeast direction. The expansion of these loops may reduce the magnetic tension above the filament F2. That is to say, the constraint of the overlying magnetic loops of the filament F2 is partly removed. The animation 3 presents the eruption of the two active-region filaments in 304 \AA\ and 131 \AA\ observations.

Due to the disturbance of the F1 eruption, the brightening appeared at the east part of the F2 at 09:21 UT (see the position marked by the blue arrow in Figs. 4(a2), 4(b2), and 4(c2)). Following the rise of  the F2, the brightening at the east part of the F2 extended toward the northeast along the direction of the F2 eruption (see the blue arrows in Figure 4(a3), 4(b3) and 4(c3)). Seen from the Figs. 1(e) and 1(f), the eastern footpoint of the F2 is rooted in the positive polarity and the negative polarity was distributed at the southeast. Therefore, the brightening was probably caused by magnetic reconnection between the F2 and the surrounding magnetic loops connecting the remote eastern positive polarity. After the eruption of the F2, the untwisting motion can be seen clearly in SDO/AIA EUV observations. Moreover, the materials of the F2 can be seen to be transferred into the large-scale magnetic loops (see the position marked by the blue arrows in Figs. 4(a4), 4(b4), and 4(c4)). The material of the F2 was transferred from the west to the east. Finally, the eastern foot-point of the F2 was found to connect the remote positive polarity (see Figs. 4(a5), 4(b5), and 4(c5)). That is to say, the magnetic structure of the F2 reconnected with the surrounding magnetic loops. Meanwhile, the F1 eruption still continued. The white arrows in Fig. 4 denote the process of whip-like eruption of the filament F1. 

The black curved dotted lines AB and CD in Figure 4(a5) outline the paths of the two filament F1 and F2 material transfer, respectively. Along the two trajectories, two time-distance diagrams were made to show the speed of the material transfer along these large-scale loops in 304 \AA\ wavelengths (see Figure 5). The velocity of the material transfer of the filament F1 along the slit AB is about 800 km$^{-1}$. The velocity of the material transfer of the filament F2 along the slit CD is about 300 km$^{-1}$. Meanwhile, the time lag between the occurrence of F1 eruption and that of F2 eruption is about five minutes.

To explain the process of the F1 and F2 eruptions, a schematic diagram is illustrated in Figure 6 to show the change of the magnetic structures of filament system before and after its eruption according to the observations of H$\alpha$ and EUV images. The full and the open patches indicate the negative and the positive polarities. The filament F1 is marked by the purple sigmoid line in panel (a). The cyan arched lines indicate the overlying magnetic loops of the filament F1. The red curve lines indicate the twisted magnetic structure of the filament, which may form during the filament F2 eruption. The blue closed loops indicate the large-scale surrounding magnetic loops before the filament eruption. The orange arched lines indicate the large-scale magnetic loops overlying the filaments F1 and F2. After the filament F1 eruption, the material of the filament F1 was injected into the overlying magnetic loops marked by the purple dots in panel (b). The green and the pink curves indicate the newly formed magnetic loops after the magnetic reconnection between the filament F2 and the surrounding magnetic loops in panels (b) and (c). 

\subsection{Filament chirality}
As this active region is located in the northern hemisphere, Chances are that it has negative helicity (Martin 1998; Pevtsov et al. 2003; Ouyang et al. 2017). However, the whirl of the super-penumbral fibrils is clockwise. It means that this active region exhibits positive helicity. That is to say, the host sunspot has an `` abnormal" helicity (i.e., - in the northern hemisphere and + in the southern hemisphere). According to the untwisting direction of the filament F2 during its eruption (clockwise rotation), it implies that its helicity is positive. Therefore, the helicity of the filament F2 is consistent with that of the active region. Generally, current helicity and magnetic helicity have the same sign. To confirm the helicity of the active region, the $\alpha$ factor is also calculated by using vector magnetograms observed by SDO/HMI. A positive and negative $\alpha$ value corresponds to positive and negative helicity. Under the force-free condition in the photosphere, the relation between $\alpha$ factor and magnetic field is as follows:

\begin{equation}\label{equation1}
\bigtriangledown \times {\bf B}= {\bf \alpha} {\bf B}.
\end{equation}

In order to calculate the $\alpha$ factor, the electric current should be known. The calculation of current is based on Ampere's law:
\begin{equation}\label{equation1}
{\bf J}=\frac{1}{\mu_0}(\bigtriangledown \times {\bf B}),
\end{equation}
in which $\mu_0$ is the vacuum magnetic permeability and ${\bf B}$ denotes the vector magnetic fields. The current density perpendicular to the solar surface can be calculated from the Ampere's law by using HMI vector magnetograms according to the following equation:
\begin{equation}\label{equation2}
{\bf J_z}=\frac{1}{\mu_0}(\bigtriangledown \times {\bf B})_z=\frac{1}{\mu_0}(\frac{\partial B_y}{\partial x}-\frac{\partial B_x}{\partial y}),
\end{equation}
where $B_x$ and $B_y$ are the two components of the  photospheric horizontal magnetic field. Finally, the $\alpha$ factor can be obtained using the following equation:

\begin{equation}\label{equation2}
{\bf \alpha}=\mu_0\frac{\bf J_z}{\bf B_z},
\end{equation}
where $\bf B_Z$ is the component of the  photospheric vertical magnetic field.

Figure 7 shows a vector magnetogram at 09:12 UT and $\alpha$ value distribution of the active region before the filament F2 eruption. The blue and the red arrows denote the transverse magnetic fields from the positive and the negative polarities, respectively. The red and the blue color in the right panel of Figure 7 denote the positive and the negative value of the $\alpha$ factor. The average $\alpha$ values of the contoured region of the leading sunspot (the magnetic field strength lower than -300 G) is 4.4 Mm$^{-1}$. The average $\alpha$ value in the black box (positive polarity) is 1.8 Mm$^{-1}$. It means the helicity of the filament F2 is the same as that of the leading sunspot and the positive polarity in the black box, in which the two foot-points of the filament F2 rooted. However, the average $\alpha$ value in the region under the main body of the filament marked by the green box is -1.3 Mm$^{-1}$. The helicity of this region is opposite to that of the two footpoints of the filament. Ouyang et al. (2015) also calculated the $\alpha$ values at the filament locations. They found that the average $\alpha$ values are consistent with the helicity sign of the two filaments. The opposite helicity of the filament location and the two foot-point regions may be caused by the weak transverse magnetic fields under the filament F2. The measure errors of the transverse magnetic field are larger in the weak magnetic field region. Note that we ignore those pixels with $\mid$$B_z$$\mid$ $\leq$ 10 G and transverse field $\leq$ 100 G to reduce the measurement error. According to the results of Martin (1998), a positive helicity should correspond to left-bearing filament thread. However, the filament F2 has right-bearing threads as indicated by Figure 2. As pointed out by Chen et al. (2014), the discrepancy implies that the filament F2 is supported by a sheared arcade. However, the obvious untwisting motion occurs after the filament eruption. These observations may imply that a flux rope formed in the filament eruption phase. 

\section{Conclusion and discussion}\label{sec:conclusion}

In this paper, we studied the process of an active region filament ( F2) eruption associated with a B-class flare, which might be triggered by a neighboring filament (F1) in AR NOAA 12740 on May 7, 2019. The filament F1 experienced a whip-like eruption and the material of the F1 was transferred into the large-scale magnetic loops. Soon, the F2 was observed to be rolling from one side to the other before its eruption and exhibited untwisting motion after its eruption according to the H$\alpha$ center and off-band observations. As the filament F2 exhibited clockwise rotation if viewed from the top to the solar surface, the filament F2 is determined to be positive helicity. Furthermore, the material of the F2 was also transferred into the large-scale magnetic loop via magnetic reconnection between the filament and the surrounding magnetic loops during its eruption.

Recently, Awasthi et al. (2019) found the rolling motion exists in an active-region filament by using the NVST off-band H$\alpha$ images. They think that the rotating motion about the filament spine strongly support the present of twisted field lines in the vicinity of the filament spine. In this study, it is also very clear that there is a rolling motion of the filament spine before its eruption. Furthermore, the obvious untwisting motion of the filament was also observed by tracing the movement of the filament threads and can be seen from the Doppler images during its eruption. 

The time interval between the onset of two filament eruptions is about five minutes. The filament F1 eruption looks like a whip-like motion with the northern part erupting and the southern end being almost fixed. The material of F1 was observed to be injected into the large-scale magnetic loops, in which one end of foot-points is rooted near the middle of F2. {Due to the filament F1 eruption, the large-scale magnetic loops expands toward the northeast, which can be seen from a series of observations in 131 \AA\ wavelength. The expansion of these loops may reduce the magnetic tension above the filament F2, which resulted in the loss of equilibrium of the filament F2.} Therefore, the filament F2 eruption is possibly caused by the disturbance of F1 eruption. The two filament eruptions have a certain physical connection like the simulation of T{\"o}r{\"o}k et al (2011), which is called a sympathetic event. 

A handedness property known as``chirality" has been discovered for filament channels and filaments by Martin, Bilimoria, \& Tracadas (1994). By using the fibril patterns, filament spines, barbs, and overlying arcades of coronal loops, Martin (1998) presents a detailed method to determine the chirality of filament channels and filaments. This method was developed by Chen et al. (2014) to judge magnetic structure of solar filaments. According to the results of Martin (1998), Pevtsov et al. (2003), and Ouyang et al. (2017), most solar filament channels and filaments in the northern (southern) hemisphere is dextral (sinistral). Chae (2000) proposed a method to determine the magnetic helicity of the filament, i.e., if the upper thread is right skewed relative to the lower one, the filament has positive helicity. On the contrary, the filament has negative helicity. According to the direction of the untwisting motion, it is determined that the filament F2 has positive helicity. From the sunspot fibrils in TiO images and the continuum intensity images, we have not found the existence of the obvious whirls. But the super-penumbral fibrils of this active region in the chromosphere have noticeable clockwise whirl, which is not the prevailing chirality in the northern hemisphere by Pevtsov, Balasubramaniam, \& Rogers (2003). 

According to the results of Martin (1998), a filament with a positive helicity should correspond to left-bearing filament threads. However, the filament F2 has right-bearing threads. As pointed out by Chen et al. (2014), the discrepancy implies that the filament F2 is supported by a sheared arcade. To confirm the helicity signal of the filament F2, the $\alpha$ values in the two footpionts and the location of the filament F2 are also calculated by using vector magnetograms. It is found that the helicity of the filament F2 is consistent with that of the two footpints and opposite to that of its location. Ouyang et al. (2015) studied the helicity of the two filaments and found that the helicity of the two filaments in their locations is consistent with that of the filaments. We think the negative helicity at the filament location may be caused by the weak transverse magnetic fields compared with the two footpoints of the filament F2. The helicity of the footpints of solar filaments may be more reliable for judging the helicity of solar filaments than that of filament location. As the obvious untwisting motion during the filament eruption, it can be deduced that the flux rope formed at the early stage of the filament eruption through magnetic reconnection. This filament F2 is located in the northern hemisphere. However, the filament F2 has not the prevailing chirality in the northern hemisphere discovered by Martin (1998) and Pevtsov et al. (2003).

Usually, the filament materials fall down along the axis of the filament to the foot-points of filaments due to the gravity during the filament eruption. In this event, the filament materials were transferred into the surrounding large-scale magnetic loops. This phenomenon is explained as the transfer of the filament materials caused by magnetic reconnection between the twisted magnetic structure of the filament and the surrounding magnetic loops. In standard solar eruption model, magnetic reconnection occurs below a filament or a flux rope and a current sheet forms under the eruptive magnetic structure. Therefore, the event studied in this paper is different from the classical model. Aulanier, \& Dud{\'\i}k (2019) have carried out the simulation on magnetic reconnection between arcade loops and a flux rope, which finally produced a flux rope and flare loops. The scenario is very similar with this event. Li et al. (2016) also reported magnetic reconnection between a quiescent filament and nearby coronal loops. Comparing with Li et al.'s event, we found more complicated motion features in our study such as rolling motion, untwisting motion, and material transfer. In addition, we suggest that a part of the twist in the filament was transferred into the surrounding magnetic loops, resulting in its failed eruption. As described above, this study extends the standard solar eruption model and sheds a light to understand complex solar eruption.

The authors thank the referee's careful reading the manuscript and constructive comments that helped to improve this paper. We would like to thank the NVST, SDO/ AIA, and SDO/ HMI teams for the high-cadence data support. This work is sponsored by the National Science Foundation of China( NSFC) under the grant numbers 11873087, 11603071,11503080, 11633008, by the Youth Innovation Promotion Association, CAS (No.2011056) , by the Yunnan Key Science Foundation of China under number 2018FA001 and Yunnan Science Foundation for Distinguished Young Scholars, by Key Laboratory of Dark Matter and Space Astronomy, CAS, by project supported by the Specialized Research Fund for State Key Laboratories and by the grant associated with project of the Group for Innovation of Yunnan Province. X.L.Y. thanks ISSI-BJ for supporting him to attend the team meeting led by J. C. Vial and P. F. Chen.

\begin{figure}
\plotone{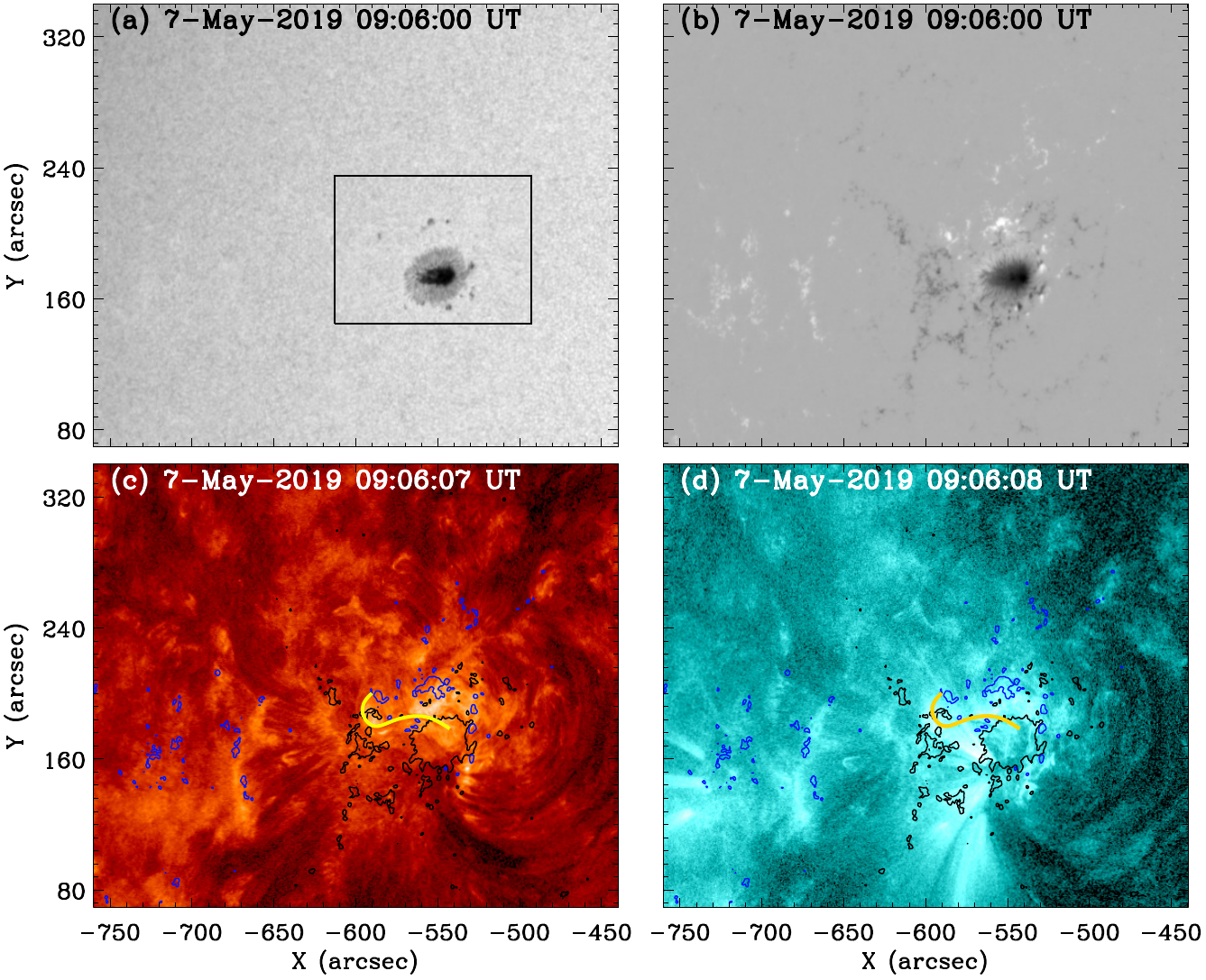}
\caption{Appearance of active region NOAA 12740. (a) and (b): Active region NOAA 12740 in the continuum image and line-of-sight magnetogram. (c) and (d): EUV observations (304 \AA\ and 131 \AA\ images) of this active region with the line-of-sight magnetogram superimposed. The levels of the magnetic contour are $\pm$ 200 G. The yellow curved lines in Figures 1(c) and 1(d) show the location of the filament F2. The black box in Figure 1(a) indicates the field of view of Figures 2 and 3.\label{fig1}}
\end{figure}

\begin{figure}
\plotone{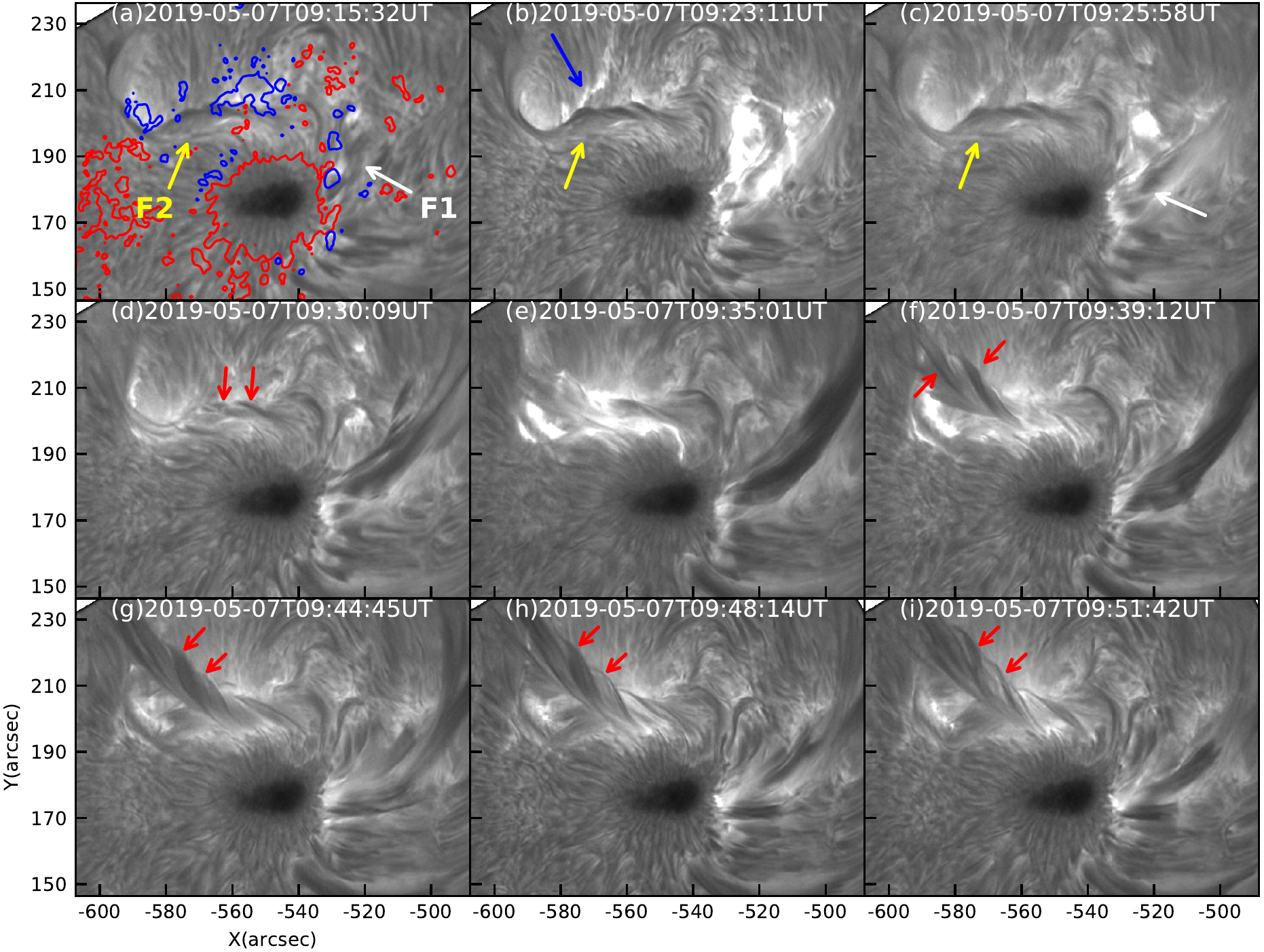}
\caption{Series of H$\alpha$ center images observed by the NVST showing the eruption of the two active-region filaments in active region NOAA 12740. H$\alpha$ image at 09:15 UT superimposed by the line-of-sight magnetogram in Figure 1(a). The contour values of magnetic fields are $\pm$ 100 G. The red patches and the blue contours denote the negative and the positive magnetic polarities, respectively. The yellow arrows in Figs. 1(a)-1(c) denote the active-region filament F2. The white arrows in Figs. 1(a) and 1(c) denote the active-region filament F1. The red arrows in Figs. 2(d), 2(f), 2(g)-2(i) point to the threads of F2. \label{fig1}}
\end{figure}

\begin{figure}
\plotone{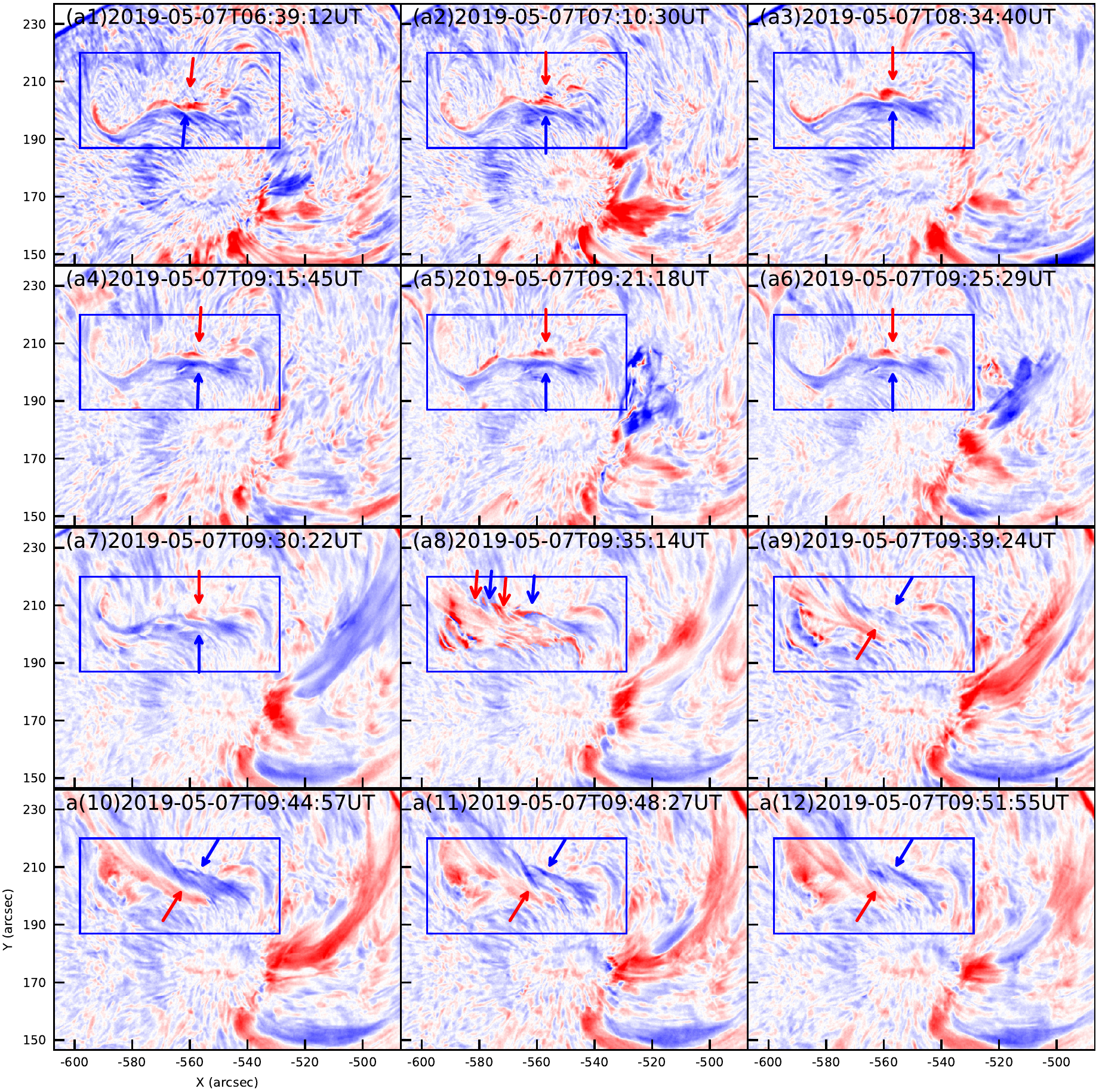}
\caption{Constructed Dopplergrams by using H$\alpha$ off-band $\pm$ 0.4 \AA\ images observed by the NVST. The spine is divided into the southern and the northern part marked by the blue and the red arrows in Figure 3(a). The blue and the red arrows denote blueshift and redshift at the opposite sides of the filament spine in Figures  3(a1)-3(a7) and Figures 3(a9)-3(a12). The blue boxes outline the location of F2. \label{fig3}}
\end{figure}

\begin{figure}
\plotone{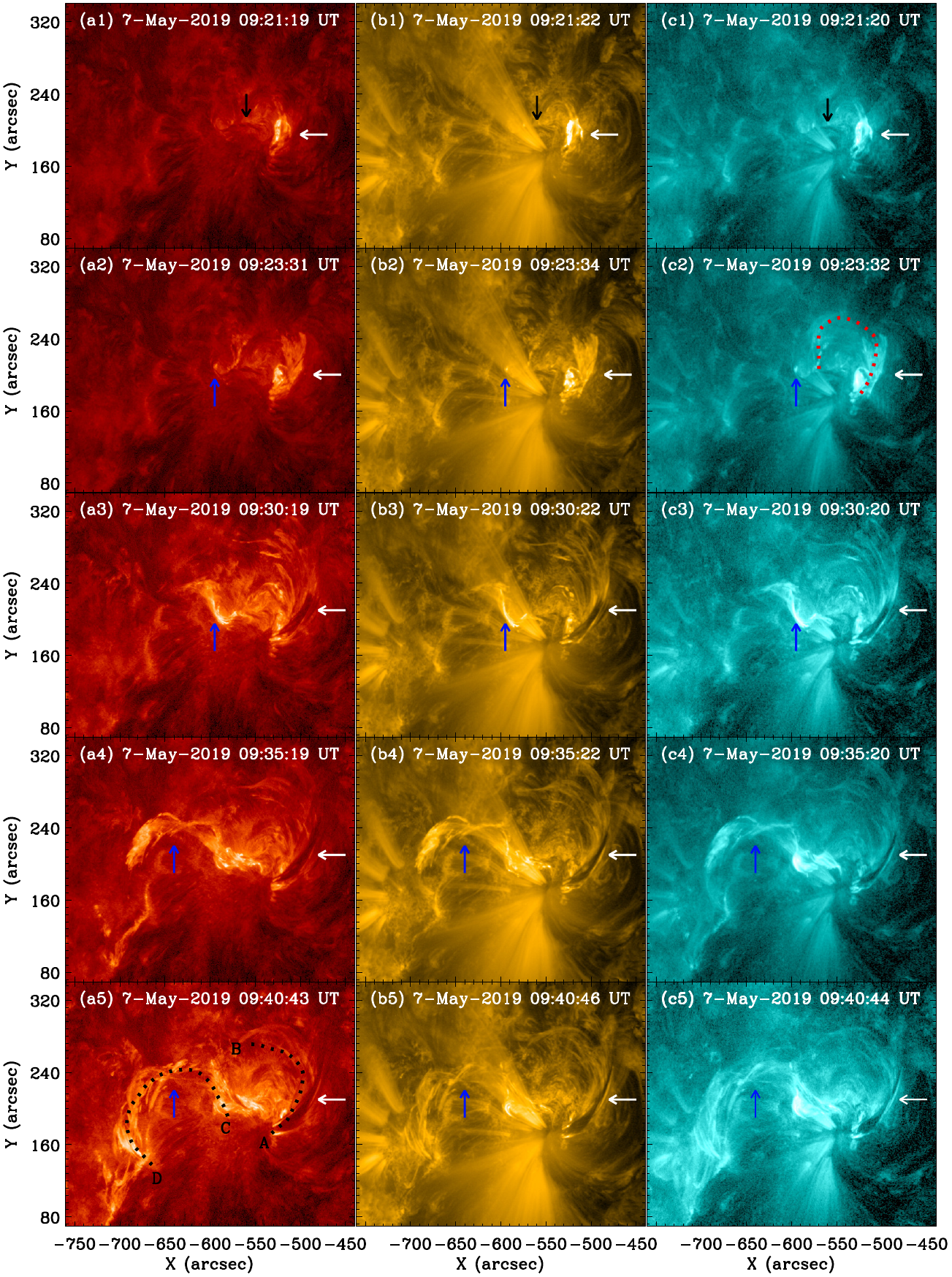}
\caption{Series of EUV images (304 \AA, 171 \AA, 131 \AA\ images) observed by SDO/AIA showing the process of two filament eruptions. The white and the black arrows in Figure 4(a1), 4(b1), and 4(c1) denote the locations of F1 and F2. The white arrows in Figures (a2)-(a5), (b2)-(b5), (c2)-(c5) denote the blowout jet produced by F1 eruption. The blue arrows in Figures (a2), (b2), and (c2) denote the brightening at the beginning of F2 eruption. The blue arrows in Figures (a3)-(a5), (b3)-(b5), and (c3)-(c5) denote the material of the filament F2. The red dotted line in Figure 4(c2) indicates the trajectory of the F1 material propagation. The black dotted lines in Figure 4(a5) denote the position of the slits AB and CD, along which the time-distance diagrams were made in Figure 5.
  \label{fig2}}
\end{figure}

\begin{figure}
\plotone{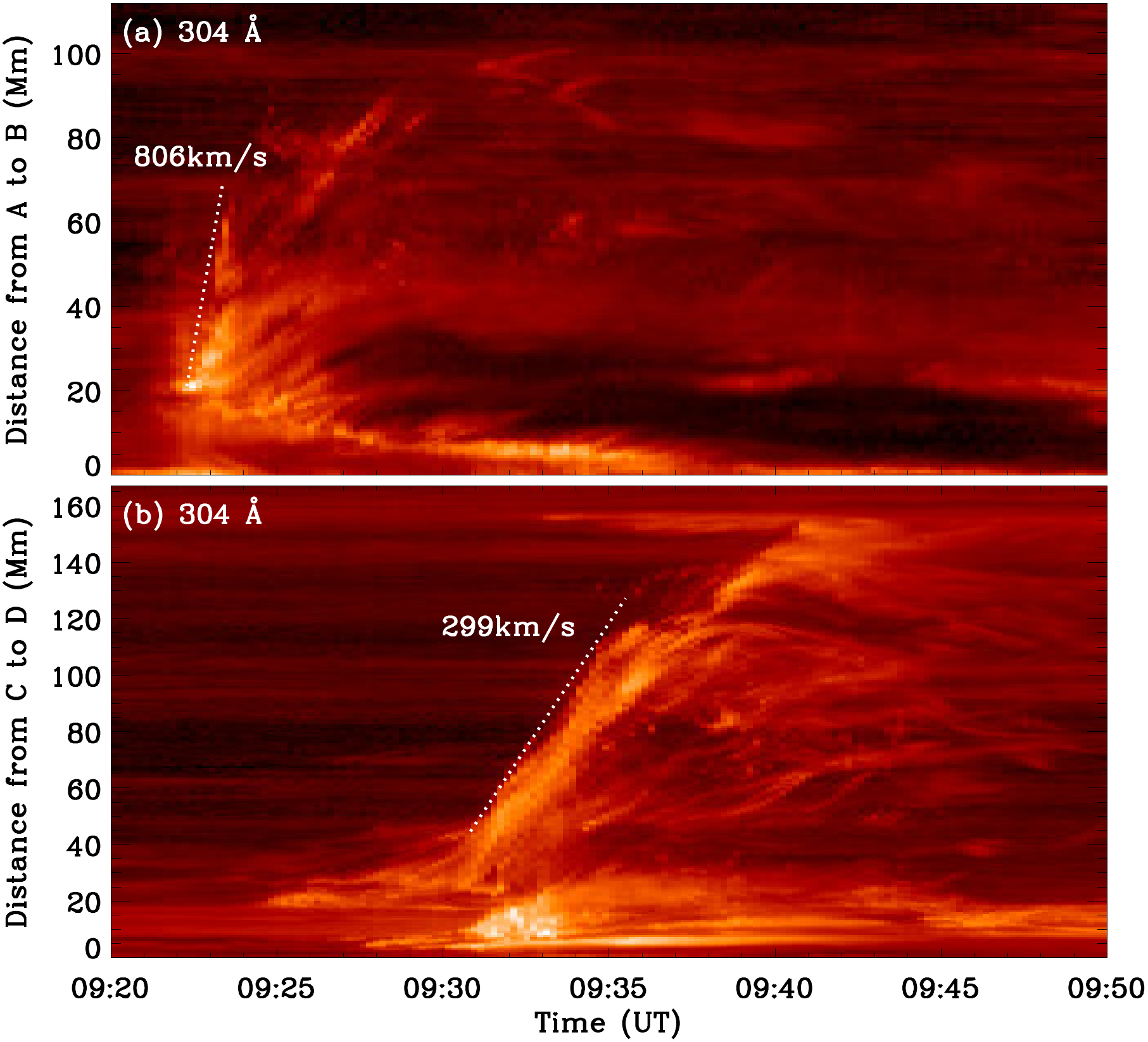}
\caption{(a): A time-distance diagram along the slits AB by using 304 \AA\ images. (b): A time-distance diagram along the slit CD by using 304 \AA\ images. The black dotted lines in Figure 4(a5) outline the position of the slits AB and CD. \label{fig6}}
\end{figure}

\begin{figure}
\centering
\includegraphics[angle=0,scale=0.4]{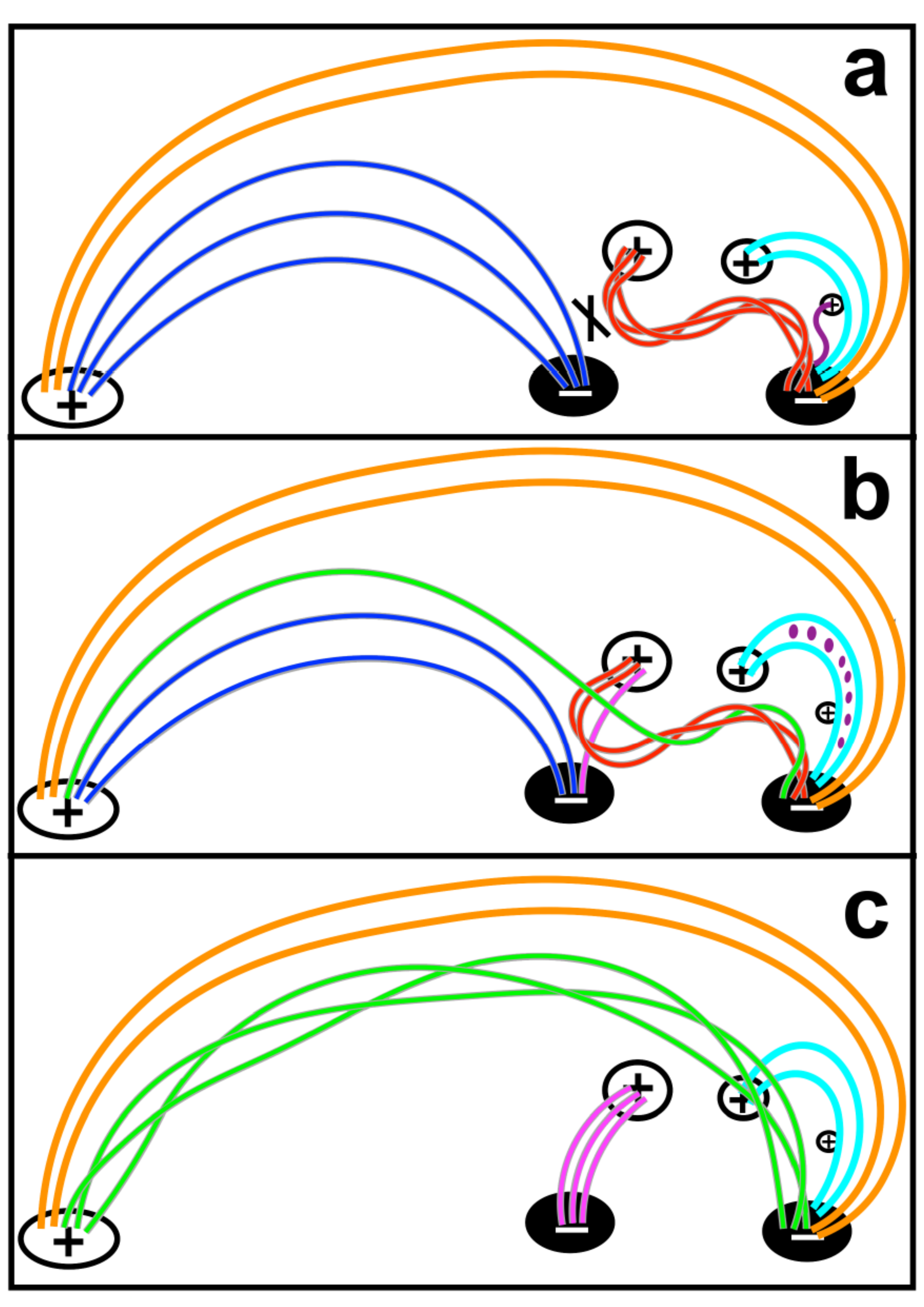}
\caption{Schematic diagram illustrating the change of the magnetic structures of the filament F2 system in its eruption. The full and the open patches indicate the negative and the positive polarities. The filament F1 is marked by the purple sigmoid line in panel (a). The cyan arched lines indicate the overlying magnetic loops of the filament F1. The red curve lines indicate the twisted magnetic structure of the filament F2. The blue closed loops indicate the large-scale surrounding magnetic fields before the filament F2 eruption in panel (a). The green and the pink curves indicate the newly formed magnetic loops after magnetic reconnection between the filament F2 and the surrounding magnetic fields. The purple dots in panel (b) represent the the material of the filament F1 that injected into the overlying magnetic loops. The orange arched lines indicate the large-scale magnetic loops overlying the filaments F1 and F2.\label{fig6}}
\end{figure}

\begin{figure}
\centering
\includegraphics[angle=0,scale=1]{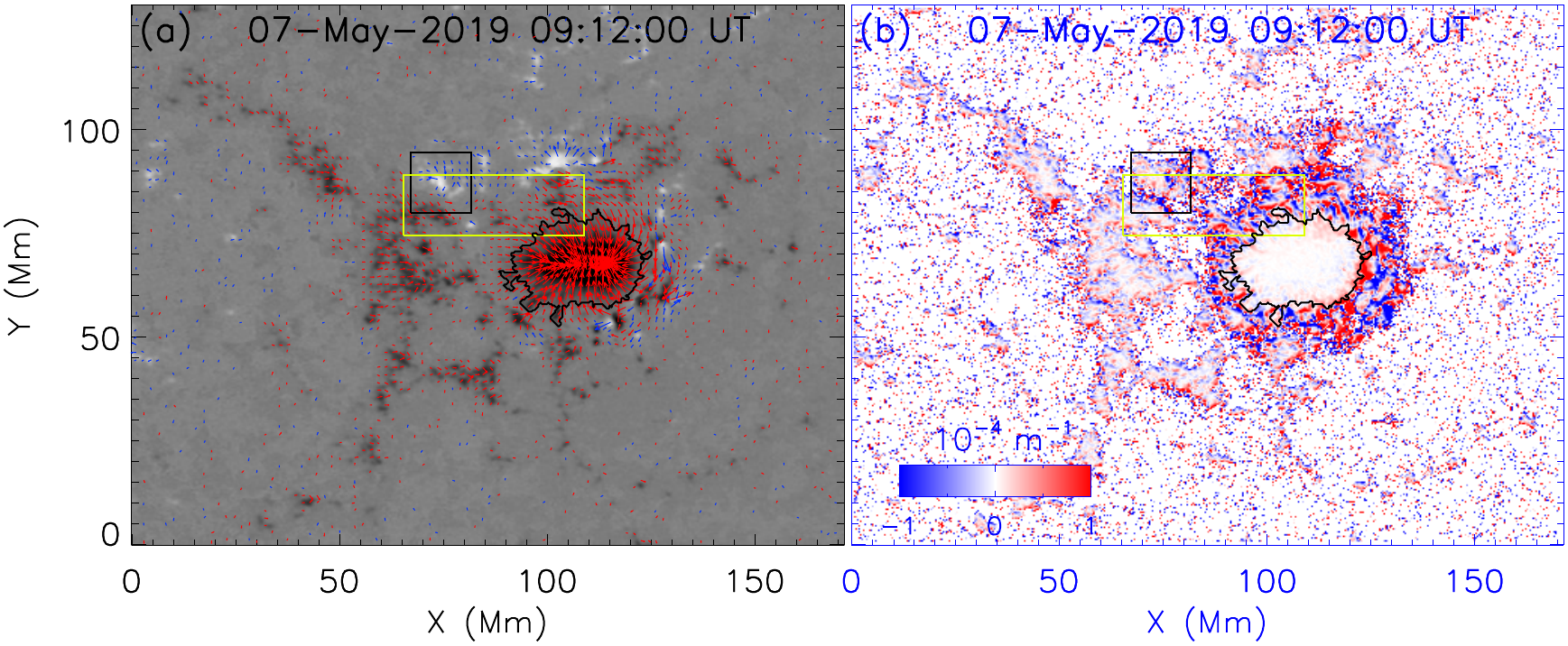}
\caption{A vector magnetogram map at 09:12 UT (left panel) and $\alpha$ value distribution of the active region (right panel) before the filament F2 eruption. The blue and the red arrows denote the transverse magnetic fields from the positive and the negative polarities in the left panel, respectively. The red and the blue color in the right panel denote positive and negative value of the $\alpha$ factor in the right panel. The regions in the black box and in the curved line are used to calculate the $\alpha$ values of the two foot-points of the filament F2. The region in the green box is used to calculate the $\alpha$ value of the filament F2 body.\label{fig6}}
\end{figure}

\end{document}